\documentclass[superscriptaddress,twocolumn,10pt, preprintnumbers,amsmath,amssymb,colorlinks,floatfix]{revtex4-2}

\usepackage{graphicx}
\usepackage{dcolumn}
\usepackage{bm}
\usepackage{array}
\usepackage{amsmath,bm}
\newcolumntype{P}[1]{>{\centering\arraybackslash}p{#1}}

\usepackage{lipsum}

\usepackage{nameref}
\usepackage{varioref}
\usepackage{hyperref}
\usepackage{cleveref}
\usepackage{comment}

\newcommand{\caruo}{Ca$_3$Ru$_2$O$_7$}
\newcommand{\AFMa}{AFM$_a$}
\newcommand{\AFMb}{AFM$_b$}

\begin{document}

\title{The magnetic structure and field dependence of the cycloid phase mediating the spin reorientation transition in \caruo}

\author{Q.~Faure}
\email[Corresponding author.~Electronic address: ]{quentin.faure@esrf.fr}
\affiliation{London Centre for Nanotechnology and Department of Physics and Astronomy,University College London, Gower Street, London WC1E6BT, UK}
\affiliation{European Synchrotron Radiation Facility, Grenoble, France}

\author{C.~D.~Dashwood}
\affiliation{London Centre for Nanotechnology and Department of Physics and Astronomy,University College London, Gower Street, London WC1E6BT, UK}

\author{C.~V.~Colin}
\affiliation{Institut Néel, Université Grenoble Alpes, CNRS, Grenoble, 38042, France}

\author{R.~D.~Johnson}
\affiliation{London Centre for Nanotechnology and Department of Physics and Astronomy,University College London, Gower Street, London WC1E6BT, UK}

\author{E.~Ressouche}
\affiliation{Univ.~Grenoble Alpes, CEA, IRIG/MEM/MDN, F-38000 Grenoble, France}

\author{G.~B.~G.~Stenning}
\affiliation{ISIS Neutron and Muon Source, STFC, Rutherford Appleton Laboratory, Didcot, OX11 0QX, United Kingdom}

\author{J.~Spratt}
\affiliation{The Natural History Museum, Imaging and Analysis Centre, Cromwell Road, London SW7 5BD, United Kingdom}

\author{D.~F.~McMorrow}
\affiliation{London Centre for Nanotechnology and Department of Physics and Astronomy,University College London, Gower Street, London WC1E6BT, UK}

\author{R.~S.~Perry}
\email[Corresponding author.~Electronic address: ]{robin.perry@ucl.ac.uk}
\affiliation{London Centre for Nanotechnology and Department of Physics and Astronomy,University College London, Gower Street, London WC1E6BT, UK}
\affiliation{ISIS Neutron and Muon Source, STFC, Rutherford Appleton Laboratory, Didcot, OX11 0QX, United Kingdom}

\date{\today}

\begin{abstract}
We report a comprehensive experimental investigation of the magnetic structure of the cycloidal phase in \caruo, which mediates the spin reorientation transition, and establishes its magnetic phase diagram. In zero applied field, single-crystal neutron diffraction data confirms the scenario deduced from an earlier resonant x-ray scattering study: between $46.7$~K $< T < $ 49.0~K the magnetic moments form  a cycloid in the $a-b$ plane with a propagation wavevector of $(\delta,0,1)$ with $\delta \simeq 0.025$ and an ordered moment of about 1~$\mu_{\rm{B}}$, with the eccentricity of the cycloid evolving with temperature. In an applied magnetic field applied parallel to the $b$-axis, the intensity of the $(\delta,0,1)$ satellite peaks decreases continuously up to about $\mu_0 H \simeq 5$~T, above which field the system becomes field polarised. Both the eccentricity of the cycloid and the wavevector increase with field, the latter suggesting an enhancement of the anti-symmetric Dzyaloshinskii–Moriya interaction via magnetostriction effects. Transitions between the various low-temperature magnetic phases have been carefully mapped out using magnetometry and resistivity. The resulting phase diagram reveals that the cycloid phase exists in a  temperature window that expands rapidly with increasing field, before transitioning to a polarised paramagnetic state at 5 T. High-field magnetoresistance measurements show that below $T\simeq 70$~K the resistivity increases continuously with decreasing temperature, indicating the inherent insulating nature at low temperatures of our high-quality, untwinned, single-crystals. We discuss our results with reference to previous reports of the magnetic phase diagram of \caruo\ that utilised samples which were more metallic and/or poly-domain.
\end{abstract}

\maketitle


\section{Introduction}
\label{sec:level1} 

Materials displaying competition between spin-orbit coupling, orbital physics and magnetic exchange have attracted much attention in the search for novel electronic and magnetic phases~\cite{Krempa2014, Rau2016}. Ruthenium based materials have been shown to play a particularly  significant role in this endeavour, for example, allowing fundamental concepts to be explored such as multiband superconductivity in Sr$_2$RuO$_4$~\cite{Ishida1998, Mackenzie2003} or Kitaev spin-liquids in $\alpha$-RuCl$_3$~\cite{Banerjee2016, Takagi2019}. 

\begin{figure}[!h]
\centering{\includegraphics[width=\linewidth]{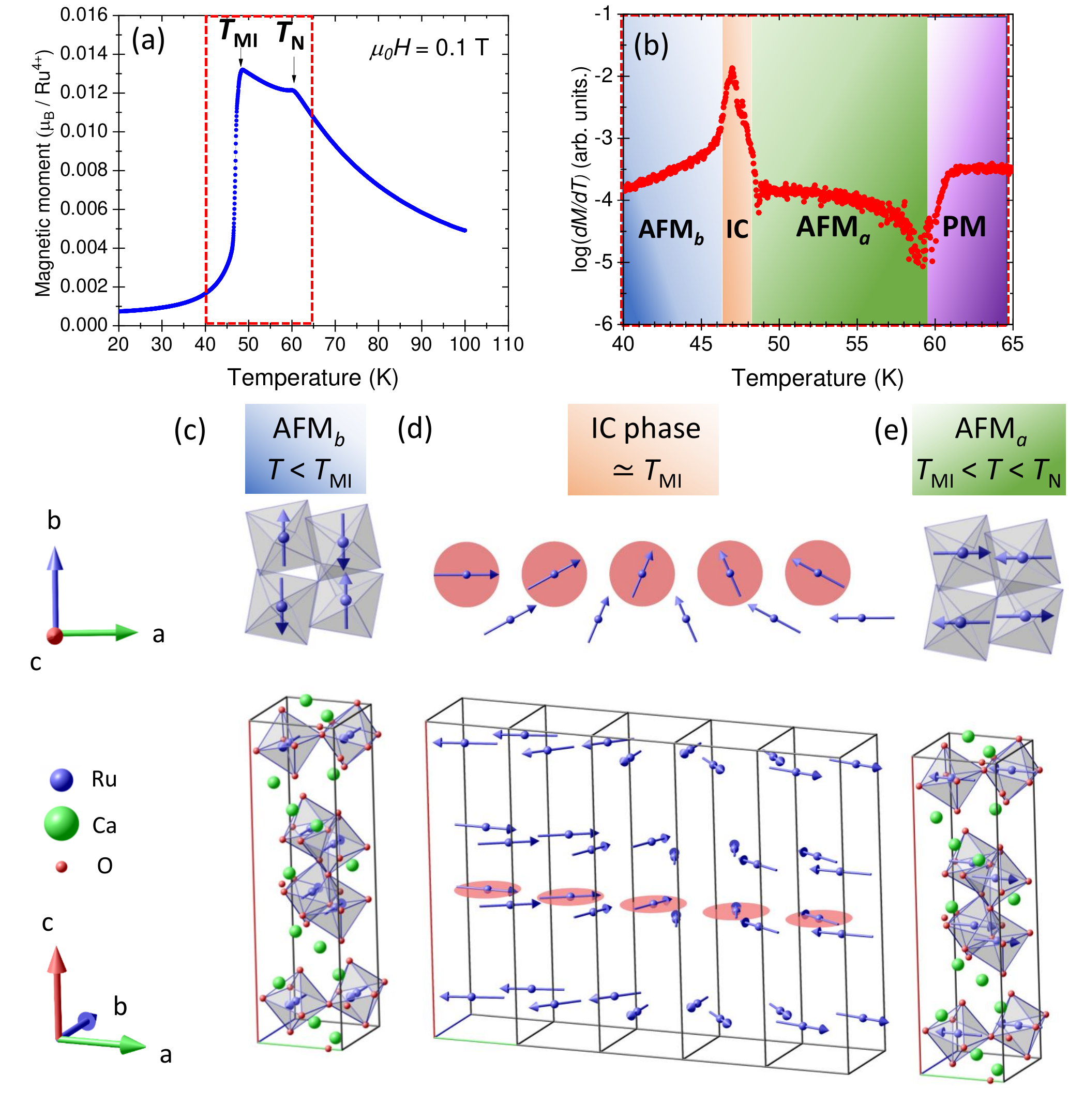}}
\caption{(a) Field cooled magnetization vs temperature at magnetic field $\mu_{0}H=0.1$~T. The N\'eel transition occurring at $T_{\rm{N}} = 60$~K and the spin reorientation transition at $T_{\rm{MI}} \simeq 48$~K are shown. The dashed red box corresponds to a smaller region of temperature depicted in (b). (b) Those two different phase transitions are better visualized through discontinuities in log(d$M$/d$T$) vs temperature. (c) Crystal and magnetic structure of \caruo\ below $T_{\rm{MI}}$ with collinear spins pointing along the $b$-axis with an antiferromagnetic arrangement between successive bilayers (\AFMb\ phase). (d) Incommensurate cycloid mediating the spin-reorientation transition around $T_{\rm{MI}}$ and defined by a propagation vector $\bm{k}_{\rm{IC}}=(\delta, 0, 1)$. Here, the incommensurate modulation is intentionally set to $\delta = 0.8$~r.l.u. for better visualization and clarity. (e) \AFMa\ phase between $T_{\rm{MI}}$ and $T_{\rm{N}}$ with spins collinear along the $a$-axis. Figures (c-e) were plotted using Mag2Pol software~\cite{Qureshi2019}.}
\label{Figure1}
\end{figure}

In this context, the Ruddlesden-Popper compound \caruo\ is also remarkable owing to the combination of magnetic ordering and polar structure that allows formation of non trivial magnetic textures~\cite{Sokolov2019, Dashwood2020, Markovi2020}. \caruo\ adopts the Bb2$_1$m space group consisting of bilayers of RuO$_6$ octahedra where tilts and rotations around the [010] and [001] directions respectively are unlocked due to the small ionic size of atoms of Ca~\cite{Yoshida2005}. Below the Néel temperature  $T_{\rm{N}} \simeq 60$~K \caruo\ becomes antiferromagnetic with the moments aligned along the $a$-axis (\AFMa\ phase) (Fig.~\ref{Figure1}e), however, the system remains metallic. Upon further cooling, it undergoes a spin-reorientation (SRT) transition around $T_{\rm{MI}} \simeq 48$~K where the moments flop from the $a$-axis to the $b$-axis (\AFMb\ phase)~\cite{Bohnenbuck2008, Bao2008} (Fig.~\ref{Figure1}c). The SRT coincides with a reconstruction of the Fermi surface linked to a metal-insulator (or low carrier density metal) transition and an abrupt change in the lattice parameters~\cite{Yoshida2005, Baumberger2006, Markovi2020, Horio2021}. 

Mediating the SRT, an incommensurate (IC) magnetic phase was recently identified~\cite{Dashwood2020}, bounded by two phases transitions at 49~K and 47~K and defined by a propagation vector of ($\delta$, 0, 1) with $\delta \simeq 0.025$ (Fig.~\ref{Figure1}d). The IC cycloid is believed to originate from the broken spatial inversion symmetry of the Bb2$_1$m space group that enables Liftshitz invariants in the free energy~\cite{Sokolov2019, Dashwood2020}. Despite the small Dzyaloshinskii-Moriya interaction energy, at the SRT, competition between the easy axis anisotropies of the AFM states suppresses the collinear order and allows the DM interaction to stabilize the IC phase.

Curiously, this IC phase has been previously observed under magnetic field~\cite{Sokolov2019} or through doping~\cite{Ke2014, Zhu2017, Lei2019} but not at zero magnetic field in undoped crystals. Instead, researchers have observed a metallic ground state below the SRT, quite different to our weakly insulating crystals\cite{Dashwood2020}. The origin of these discrepancies probably lie in extreme sensitivity of the lattice, electronic and magnetic degrees of freedom to perturbation; minute variations in stoichiometry driving macroscopic changes in the properties. Moreover, it has been demonstrated that sample twinning due to orthorhombic space group is prevalent in as-grown crystals\cite{Markovi2020}. This potentially leads to ambiguities when interpretating data taken on twinned samples, for example reference~\cite{Bao2008}. 

The motivation of this work is to revisit the $H-T$ phase diagram in untwinned crystals to map out and characterise the various ordered magnetic states. Towards this, we present magnetisation, elastic neutron scattering and resistivity measurements as a function of temperature and magnetic field aligned along the crystallographic \textit{b}-axis.

\section{Experimental methods}
\label{sec:level2}
Single crystals were grown by floating zone method using a four mirror furnace from Crystal Systems Inc. RuO$_2$ is known to be volatile at the melting point so feed rods of stoichiometry Ca$_3$Ru$_{2.5}$O$_{8.1}$ were prepared and the \textit{flux feeding} method was utilised~\cite{PERRY2004}. The total gas pressure inside the furnace was 10 bar with P$_{\rm{O}2}$=1 bar and the growth speed was 8 mm/hr. The cation composition of the as-grown crystals were confirmed by WDX EPMA; the Ca:Ru ratio was 1.48(3) and no foreign elements above oxygen were detected down to approximately 500 ppm. The oxygen content could not be reliably determined. Many of the as-grown crystals were twinned; the twin boundaries were along [110] indicating that there are orthorhombic mirror planes~\cite{Markovi2020}. Untwinned samples were identified by linear polarised light microscopy in reflection mode and cut out using a wiresaw. The neutron scattering studies confirmed that samples were a single twin. Identification and isolation of a single twin is particularly important in bulk measurements, as contributions from two different orientations of the crystallographic axes to the applied magnetic field will be averaged. \\

Measurements of the magnetisation and resistivity were made using a quantum design MPMS3 and Dynacool PPMS, respectively. Single domain samples were aligned to the \textit{b}-axis using a Laue camera to within one degree and cut by wiresaw. Four terminal measurements were used for the resistivity measurements with contacts made to the crystal via Dupont 6838 high temperature curing silver paint. Care was taken to ensure the current paths were parallel to the \textit{b}-axis and contained no \textit{c}-axis contribution. A low frequency \textit{dc} resistance measurement was used with a 1 Hz, 2 mA current.\\

The neutron diffraction experiment was performed on the CEA-CRG D23 single-crystal two-axis diffractometer with a lifting arm detector at the Institut Laue Langevin. The 30 mg oriented sample was placed in a 6~T cryomagnet so that $H \parallel b$ with less than $2^{\circ}$ of misalignment. Due to the small mass of the crystal, a wavelength of $2.38~\rm{\AA}$ from pyrolytic graphite monochromator was chosen to maximize the flux. Hence only the $h0l$ plane was accessible (up to $h = 2$ and $l = 7$).  

\section{Magnetic structure around $T_{\rm{MI}}$ at $H = 0$ (IC phase)}
\label{sec: level2bis}

We first report the results of our investigation using neutron diffraction in zero applied field of the magnetic structure of the incommensurate phase previously discovered  by a resonant x-ray scattering study~\cite{Dashwood2020}. The crystallographic and magnetic structures were refined by least-squares method using the FULLPROF software~\cite{Carvajal1993}. We begin with refining the crystallographic structure. In our high field magnet configuration, this was difficult due the restricted geometry allowing only few accessible reflections (only $h 0 l$ reflections, see Sec.~\ref{sec:level2}). As the crystal structure of \caruo\ has been previously determined and the space group remains unchanged at all temperatures, we chose to fix all atomic coordinates with published values found in Ref.~\cite{Yoshida2005}. Only the scale factor and extinction parameters were refined. 70 nuclear reflections (reducing to 21 independent ones) were collected at $T = 48.7$~K and $\mu_0H = 0$. The refined crystallographic structure is found with an agreement $R$-factor $R = 6.5 \%$ (see Appendix~A), confirming that our crystals are structurally similar to other groups'. Similar results (not shown) were extracted at other temperatures. 

Our single-crystal neutron diffraction experiments confirm that incommensurate magnetic Bragg reflections characteristic of the ICM cyclodial phase appear only in a narrow temperature region between 46.7 K and 49.0 K ($T_{\rm{MI}}$) and can be indexed by an incommensurate propagation vector $\bm{k}_{\rm{IC}} = (\delta, 0, 1)$. The incommensurate modulation $\delta$ was determined at zero field and $T= 48.7$~K through $h$-scans around $\bm{Q} = (h, 0, 1)$ (similar to those in Fig.~\ref{Figure4}a) and was found as $\delta \simeq 0.025$~r.l.u.

\begin{figure}[!h]
\centering{\includegraphics[width=\linewidth]{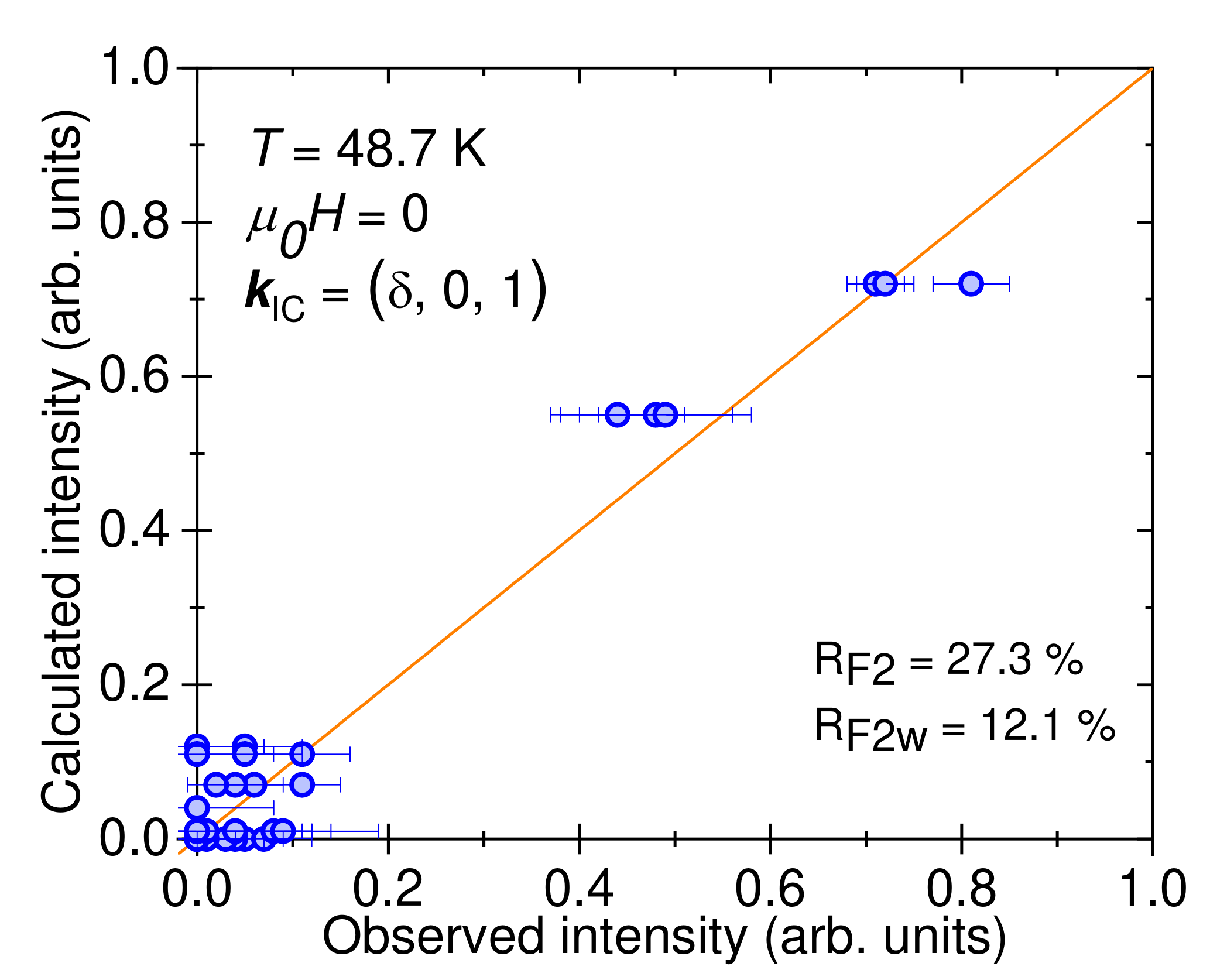}}
\caption{Representation of the magnetic structure refinement at zero field and $T = 48.7$~K performed with 59 magnetic reflections (15 independent ones) and $\bm{k}_{\rm{IC}} = (0.025$, 0, 1).}
\label{Figure2bis}
\end{figure}

Based on our previous analysis~\cite{Dashwood2020}, we model and refine the IC phase as follows (see Appendix~A for more details). For a given magnetic propagation vector $\bm{k}$ (and the associated $-\bm{k}$):
\\
$m^{j}_{l} = m^{j}_{a}\cos[2\pi(\bm{k}\cdot\bm{R_l}+\Phi_j)]\bm{x}+m^{j}_{b}\sin[2\pi(\bm{k}\cdot\bm{R_l}+\Phi_j)]\bm{y}$
\\
where $m^{j}_{l}$ is the magnetic moment of the atom j in the unit cell $l$, $m^{j}_{a}$ and $m^{j}_{b}$ the components along $a$ and $b$ of the magnetic moment, $\bm{R_l}$ is the vector joining the arbitrary origin to the origin of unit cell $l$, $\Phi_j$ is a magnetic phase, $\bm{x}$ and $\bm{y}$ the unitary vectors along $a$ and $b$ respectively. 

The refined magnetic structure is an elliptical cycloid propagating along the $a$-direction. The layers of ferromagnetically coupled magnetic moments turn in the $(a, b)$ plane and are arranged antiferromagnetically along $c$. Fig.~\ref{Figure2bis} shows the best result obtained for the refinement of the IC phase with agreement $R$-factor of $27.3 \%$. The quality of the fit is reasonable given the limited number of points available in the experiment. This result is consistent with our previous studies using resonant elastic X-rays scattering~\cite{Dashwood2020}. For completeness, refinements of the \AFMa\ and \AFMb\ phases as can be found in the Appendix~A. 

\section{Magnetic field versus temperature phase diagram}
\label{sec:level3}

We now turn to the determination of the $H - T$ phase diagram of \caruo\ around $T_{\rm{MI}}$. To that end, we used a combination of magnetometry and neutron diffraction measurements.

\subsection{Magnetometry measurements}
\label{subsec: 3: 1} 

The phase boundaries at low fields between the paramagnetic (PM), \AFMa, IC and \AFMb\ phases were determined by temperature sweep data of the magnetisation shown in Fig.~\ref{Figure1}a. We observe a clear kink in $M(T)$ at 60 K on cooling from high temperature; this is the N\'eel temperature for the \AFMa\ phase. The phase boundaries of the IC phase can be more clearly observed in the d$M$/d$T$ shown in Fig.~\ref{Figure1}b: a narrow temperature region between 49 K and 46.7~K. Below 46.7~K, the system orders in the \AFMb\ phase. This data is consistent with our previous studies~\cite{Dashwood2020}.

\begin{figure}[h!]
\centering{\includegraphics[width=\linewidth]{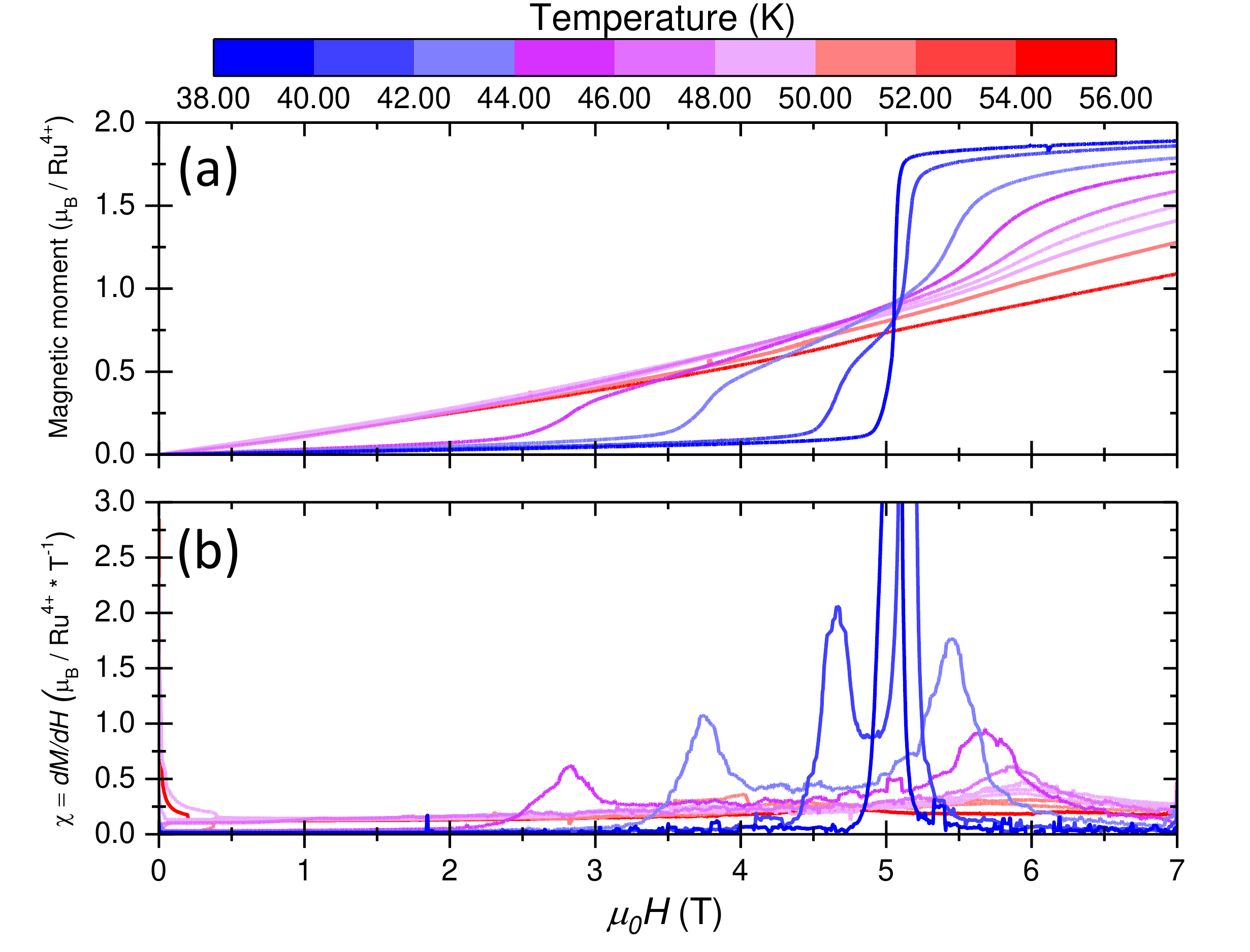}}
\caption{(a) Isothermal field dependence of magnetization curves $M(H)$ around the spin-reorientation transition for $\bm{H} \parallel \bm{b}$. (b) Isothermal field dependence of magnetic susceptibility $\chi(H)$ obtained from the first derivative of $M(H)$.
}
\label{Figure2}
\end{figure}

To map out the field temperature phase diagram, we present isothermal field sweeps of the magnetisation, \textit{M(H)} shown in Fig.~\ref{Figure2}.

At low fields, the \textit{M(H)} curves are linear as expected for a collinear antiferromagnet when the field is applied perpendicular to the easy-axis. Superlinear increases in \textit{M(H)} are observed, demonstrating the metamagnetic transitions noted in several studies~\cite{Sokolov2019, Dashwood2020,Kikugawa2021}. These transitions correspond to changes in the magnetic structure from AFM order to IC order to polarised paramagnetic state. The moment at 7 T and low temperature corresponds to 1.89~$\mu_{\rm{B}}$, slightly reduced from the expected full S=1 moment although the magnetisation has not saturated by our maximum field. About the main metamagnetic transition at 5 T are several extra transitions related to the incommensurate and commensurate magnetic phases; these are best observed in the static susceptibility shown in Fig.~\ref{Figure2}b where the peaks correspond to phase transitions. These data are collected and summarised in the temperature-field phase diagrams shown in Fig.~\ref{Figure5}, along with neutron scattering data presented below. The round blue points are the peaks in the static susceptibility. The IC phase is clearly bounded by phase transitions in both temperature and field, as we would expect for a thermodynamically distinct phase. From hysteresis in the \textit{M(H)} curves (not shown here), the low field transitions are confirmed to be first order~\cite{Kikugawa2021}. No hysteresis is observed in the the high field transition preventing clear determination of its nature. 

\subsection{Neutron scattering measurements}
\label{subsec: 3: 2}

To confirm the microscopic nature of the various ordered phases and elucidate the evolution of the IC phase under magnetic field, we performed a neutron diffraction experiment. We first examine the temperature and magnetic field dependence of the different phases extracted from temperature and field sweep scans.

\onecolumngrid

\begin{center}
\begin{figure}[b!] 
\includegraphics[width=0.92\textwidth]{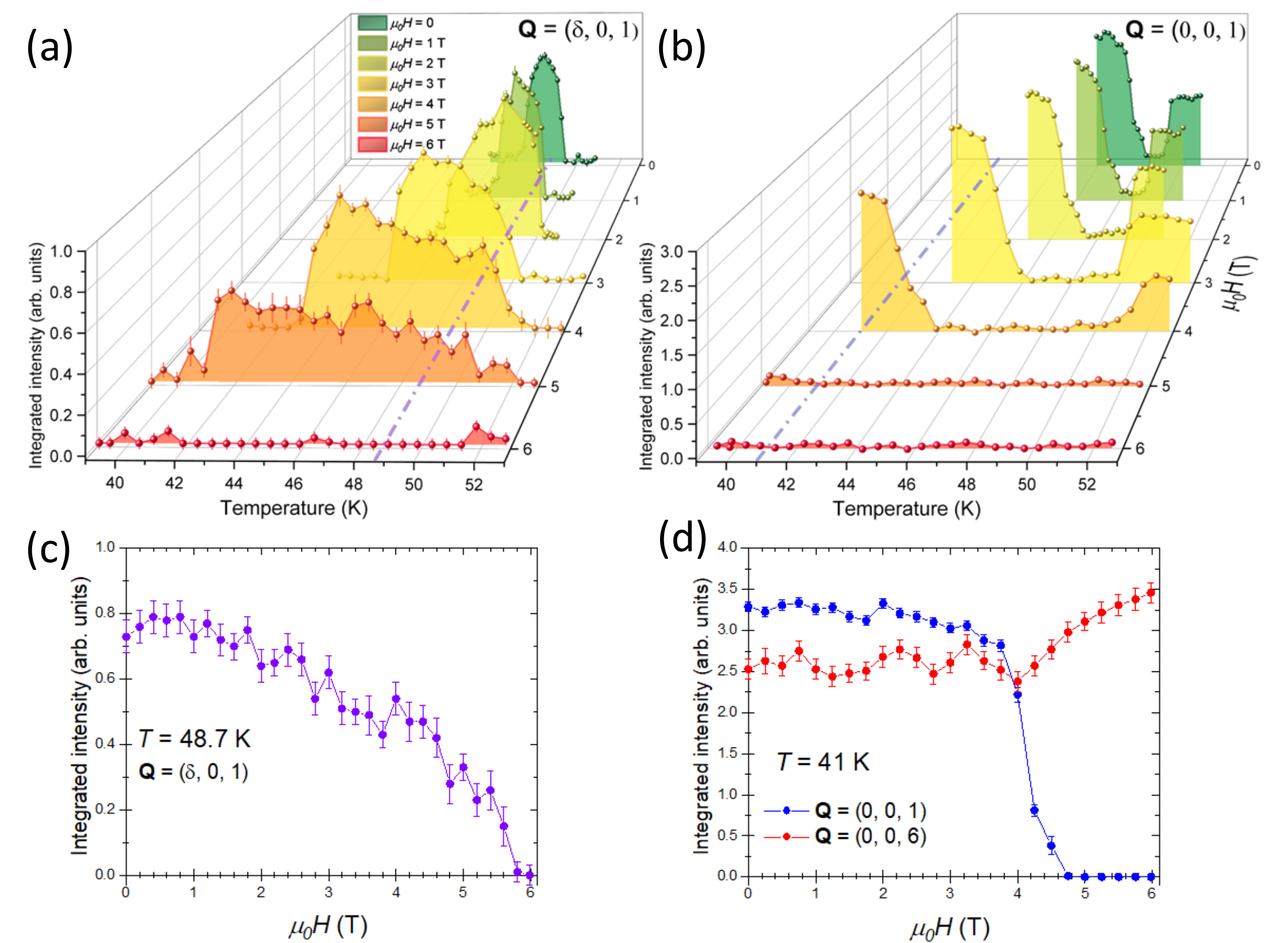}
\caption{(a-b) Temperature evolution of incommensurate and commensurate magnetic Bragg peaks $\bm{Q} = (\delta, 0, 1)$ and $\bm{Q} = (0, 0, 1)$ respectively and for different magnetic fields around the spin-reorientation transition. Purple and blue dashed-dotted lines denote field sweep measurements shown in (c-d). (c) Magnetic field evolution of incommensurate magnetic Bragg peak $\bm{Q} = (\delta, 0, 1)$ at $T = 48.7$~K. (d) Magnetic field dependence of commensurate Bragg peaks at $T = 41$~K. Blue (resp. red) points corresponds to the commensurate Bragg peak $\bm{Q} = (0, 0, 1)$ (resp. $\bm{Q} = (0, 0, 6)$) relative to the \AFMb\ phase (resp. FM phase).}
\label{Figure3}
\end{figure}
\end{center}

\twocolumngrid

Fig.~\ref{Figure3}a shows the temperature evolution of the integrated intensity obtained from rocking-curves of the incommensurate satellite $\bm{Q} = (\delta, 0, 1)$ for different values of magnetic field. At zero field, this satellite survives in a very narrow region of temperature ($\simeq 46.7 - 49.0$~K), confirming the magnetometry measurements and earlier scattering studies~\cite{Dashwood2020}. With increasing the magnetic field, this temperature region widens while the overall integrated intensity decreases and vanishes around $\mu_0H = 6$~T. In contrast, Fig.~\ref{Figure3}b shows the temperature evolution of the the magnetic Bragg reflection $\bm{Q}= (0, 0, 1)$ describing the \AFMa\ and \AFMb\ phase. Concomitantly with respect to Fig.~\ref{Figure3}a, no intensity of the commensurate phases is observed within the narrow region of the SRT while intensity is observed outside indicating phase transitions to the \AFMb\ and \AFMa\ phases. This absence of magnetic signal widens in temperature range with increasing magnetic field up to $\mu_0H = 5$~T. No intensity is found above in the whole temperature range suggesting a phase transition to a high field phase. To confirm this hypothesis, we show the magnetic field dependence of the incommensurate satellite $\bm{Q}=(\delta, 0, 1)$ measured at $T = 48.7$~K in Fig.~\ref{Figure3}c. The intensity decreases smoothly with increasing magnetic field and vanishes around $\mu_0H \simeq 5.8$~T, marking a phase transition from the IC phase to the high field phase. 

Fig.~\ref{Figure3}d shows the magnetic field dependence of both commensurate magnetic $\bm{Q}=(0, 0, 1)$ and structural $\bm{Q} = (0, 0, 6)$ Bragg peaks, related to the \AFMb\ and nuclear structure with a ferromagnetic (FM) component ($\bm{k}_{\rm{FM}}=\bm{0}$) respectively, measured at $T = 41$~K. Both intensities keep roughly constant up to $\mu_0H \simeq 4$~T. By further increasing the magnetic field, the magnetic signal of $\bm{Q}=(0, 0, 1)$ rapidly decreases to completely disappear around $\mu_0H \simeq 4.5$~T while the intensity of $\bm{Q}=(0, 0, 6)$ increases corresponding to the system acquiring a uniform magnetization. This is consistent with a transition from the \AFMb\ to a field-polarized ferromagnetic (FM) phase and also in agreement with magnetometry measurements where the saturation of the magnetic moments is around $\mu_0H \simeq 7$~T (see Sec.~\ref{subsec: 3: 1}). 
\\

\begin{figure}[!h]
\centering{\includegraphics[width=\linewidth]{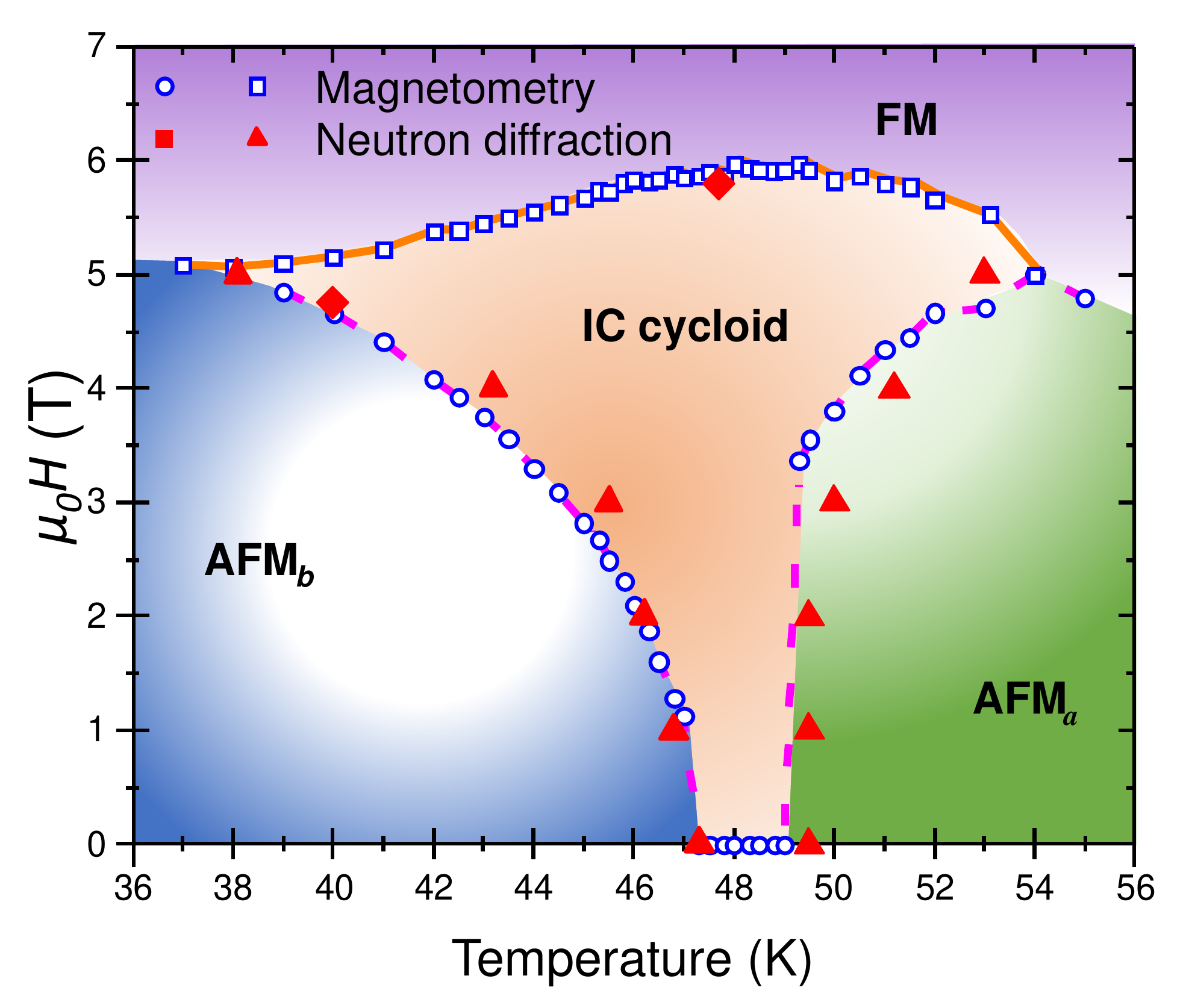}}
\caption{Magnetic field vs Temperature phase diagram of \caruo\ probed by magnetometry (empty blue points and squares) and neutron scattering (red diamonds and triangles) experiments around the spin-reorientation transition. Empty blue circles and red diamonds (resp. empty blue squares and red triangles) denote phase transitions determined from temperature sweep at fixed magnetic field (resp. magnetic field sweep at fixed temperature). The dashed pink line denotes first order transition from \AFMb\ to IC phases and from IC to \AFMa\ phases while the solid orange line denotes second order transition from the IC phase to FM phase.
}
\label{Figure5}
\end{figure}

By combining magnetometry and neutron diffraction measurements, we can now establish the $H-T$ phase diagram for our weakly-insulating single-domain \caruo\ crystals under magnetic field along the $b$-axis depicted in Fig.~\ref{Figure5}. At zero-field, the IC phase is bounded between the \AFMa\ and \AFMb\ phases in a narrow region of temperature ($\simeq 46.7-49.0$~K). The temperature region expands with increasing magnetic field up to $\mu_0H\simeq 5$~T where the IC phase disappears and is replaced with a field-polarized ferromagnetic phase.

\section{\label{subsec: 4:1} Temperature and magnetic field evolution of the incommensurate cycloid}

\begin{figure*}[htb]
\begin{minipage}[t!]{\linewidth}
\begin{center}
\centering{\includegraphics[width=\linewidth]{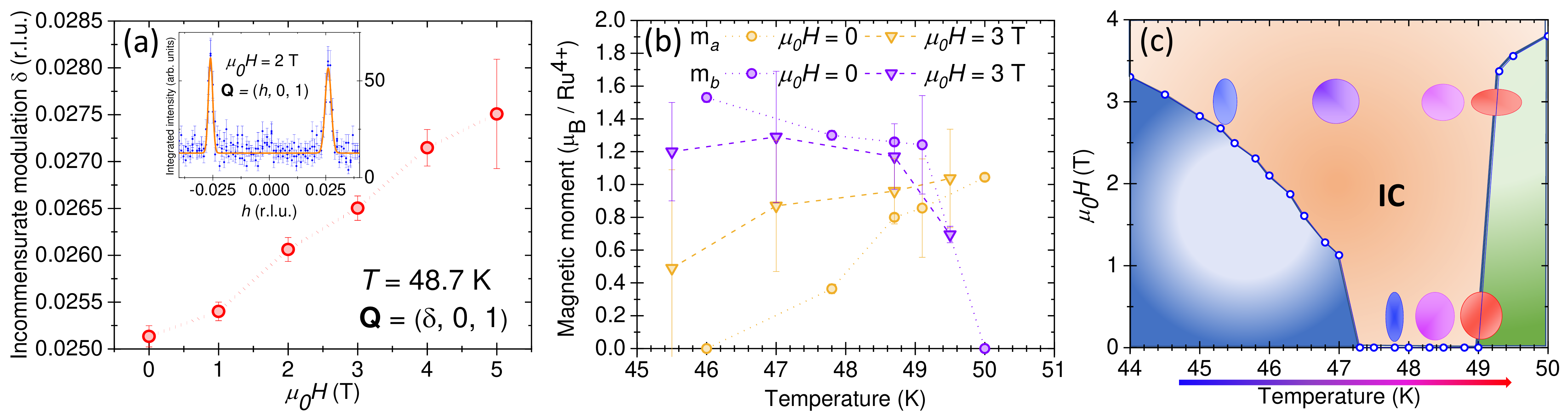}}
\caption{(a) Magnetic field dependence of the incommensurability $\delta$ (red points) at $T = 48.7$~K. The dashed red line is a guide to the eye. The inset shows a $h$-scan probed for $\bm{Q} = (h, 0, 1)$ at $T = 48.7$~K and $\mu_0H = 2$~T. The orange line is a fit constituted of two Lorentzian functions. (b) Temperature dependence of the two components $m_a$ (orange open circles and triangles) and $m_b$ (purple circles and triangles) determined from refinements (see Appendix~A) for $\mu_0H = 0$ and $\mu_0H = 3$~T. The dashed and dotted lines are guides to the eye. (c) Sketch of the evolution of the ellipsoid shape of the IC phase across the $H-T$ phase diagram.  
}
\label{Figure4}
\end{center}
\end{minipage}
\end{figure*} 

We now turn to the temperature and field dependence of the ellipsoidal moments in the IC phase. We first measured the magnetic field dependence of the incommensurate modulation $\delta$ at $T = 48.7$~K by probing the two satellites $(\pm \delta, 0, 1)$ through $h$-scans around $\bm{Q} = (h, 0, 1)$ (Fig.~\ref{Figure4}a). One can see that $\delta$ increases by 10\% with increasing the magnetic field from 0 to 5 T. 

We then repeat the procedure explained in Sec.~\ref{sec:level3} by probing 59 magnetic reflections (reducing to 15 independent ones) for various temperature and two values of the magnetic field $\mu_0H=0$ and $\mu_0H=3$~T. As the incommensurate modulation vector component $\delta$ is temperature independent, refinements were performed by fixing $\delta$ previously determined at the relevant applied magnetic field (see Fig.~\ref{Figure4}a). 
The results of the refinement are shown in Fig.~\ref{Figure4}b and Fig.~\ref{Figure4}c, describing the temperature and magnetic field evolution of the IC cycloid. Fig.~\ref{Figure4}b shows the temperature evolution of $m_a$ and $m_b$ for $\mu_0H = 0$ and $\mu_0H = 3$~T obtained from the magnetic structure refinements (see Appendix~A). The evolution of the ellipsoidal shape of the IC phase across the $H-T$ phase diagram is more clearly depicted in Fig.~\ref{Figure4}c. As previously determined by REXS in Ref.~\cite{Dashwood2020}, the eccentricity of the ellipsoid changes from elongated along $a$ to circular to elongated along $b$ with decreasing the temperature. 
At constant temperature, the magnetic field appears to have a small effect on the eccentricity of the ellipsoid at constant temperature to within the uncertainty of the measurement. However, in applied field (3 T), the eccentricity has a stronger dependence on temperature compared to zero field owing to the larger temperature window of the IC phase.
This broadening of the temperature window is likely due to the low field phase boundary is moving to lower temperatures due to the applied field along \textit{b} favouring the \AFMb\ phase. 

\section{Resistivity measurements}
\label{sec:level5} 
\begin{figure}[!h]
\centering{\includegraphics[width=0.8\linewidth]{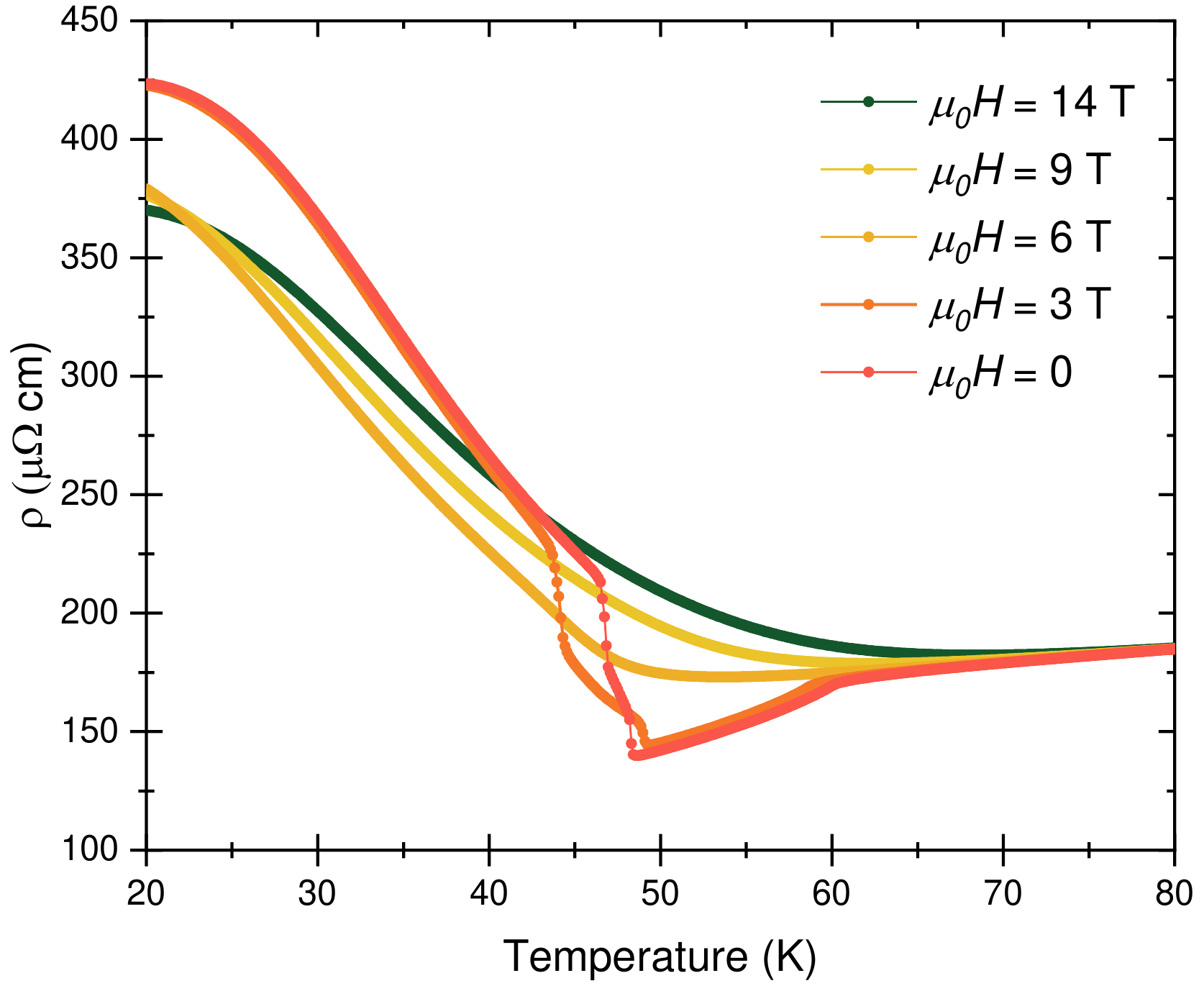}}
\caption{Resistivity measurements as function of temperature for different values of magnetic field.}
\label{Figure6}
\end{figure}

Finally, to illuminate the low energy electronic quasiparticle states, Fig.~\ref{Figure6}  shows the resistivity versus temperature at fixed magnetic fields along the \textit{b}-axis. 

At low fields, multiple features are observed at 47 K, 49 K and 60 K concomitant with the entry to the IC, \AFMa\ and paramagnetic phases with increasing temperature. As we discussed in the introduction, the ground state of our crystals is weakly-insulating. Over forty untwinned samples were measured across ten batches to check for reproducibility; all samples were weakly insulating with residual resistivities greater than 250 $\mu \Omega$cm and displayed transitions at similar temperatures. On heating from the ground state, the entry to the IC and the \AFMa\ phases are marked by discontinuous drops in the resistance, related to the proposed Fermi surface reconstruction~\cite{Markovi2020}. For comparison, high field resistance sweeps are also shown where the poorly insulating, polarised paramagnetic state is stabilised up to 70 K, with no evidence of a magnetic phase transition. This suggests that at high magnetic fields the electronic instability is independent of the long range ordered magnetic states. This is consistent with supposition that the orientation of the local moment relative to the lattice is important to the Fermi surface  reconstruction\cite{Markovi2020}.

\section{Discussion}
\label{sec:level5} 

The present study has allowed us to unveil the $H-T$ phase diagram of our weakly-insulating, single-domain crystals of \caruo\ with magnetic field along the \textit{b}-axis. Interestingly, there are discrepancies between our neutron scattering study and previous studies~\cite{Bao2008}. In Ref.~\cite{Bao2008}, the $H-T$ phase diagram with magnetic field along the \textit{b}-axis is more complicated than ours, with a field induced transition to a canted antiferromagnetic state above 5 T and below 48 K. We observe no evidence for such a state; no commensurate wave-vector or transition in the magnetisation at high fields is detected. This additional phase in the $H-T$ phase diagram is most probably due to twinned crystals misleading the interpretation of the data; the superposition of the signal for $\bm{H} \parallel \bm{b}$ and $\bm{H} \parallel \bm{a}$ would explain those observations. To check this hypothesis, we probed the $H-T$ phase diagram with $\bm{H} \parallel \bm{a}$ (see Appendix~B). The steep phase transition line around 48~K and 4~T is remarkably similar to the one found in Ref.~\cite{Bao2008} leading us to conclude that their crystals were twinned.

More recently, Sokolov \textit{et al.} observed a \textit{metamagnetic texture} by Small-Angle Neutron Scattering(SANS) that was qualitatively similar to the IC phase~\cite{Sokolov2019}. However, this phase emerged above 2 T around 48 K but was not observed in zero field, supported by magnetocaloric measurements ~\cite{Kikugawa2021}. The observed incommensurate wave-vector was around 0.06 r.l.u. and is likely to be a higher harmonic of the cycloid state that characterises field induced anharmonicity of the cycloid described by higher order terms in the free energy (see Supplementary Material in Ref.~\cite{Dashwood2020}). Hence, our previous experiment~\cite{Dashwood2020} along with the present results are consistent with Sokolov's results; the proposed \textit{improper Dzyaloshinskii textures} are higher harmonics of the cycloid phase that we observe in zero field. 

Interestingly, we observe a clear increase in the incommensurate wave vector as a function of magnetic field along the $b$-axis, with around a $10 \%$ increase in $\delta$ between 0 T and 5 T. This can be qualitatively understood within the following argument. The IC state is stabilised by effective suppression of single ion anisotropy by competing single ion terms (i.e. moment along $a$ or $b$). This allows the energy scale of the inherent DM interaction in the polar structure to become relevant, stabilising the IC order. The wavevector is related to the ratio between energy scales of antisymmetric DM and symmetric exchange. Hence, the observed increase in $\delta$ with increasing field can be attributed to an enhancement of the DM interaction. The mechanism might be due to magnetostriction, with the polar distortions of the lattice increasing in applied field. 

We now turn our attention to the nature of the transition. The microscopic origin of the spin reorientation, metal-insulator transition is still under debate. Recent work by Markovic and co-workers have proposed that a Rashba-like coupling between the local spin moments and the polar lattice distortion is linked to the Fermi surface reconstruction~\cite{Markovi2020}. 
Essentially, aligning the moments to the $b$-axis via the spin-flop transition enables the Rashba-like coupling, hybridising the bands at the Fermi level causing the gapping of low energy states. 
Above the spin-flip transition at $\sim$5 T, the application of a magnetic field along the \textit{b}-axis polarises the moments along $b$ in the paramagnetic phase. 
Within Markovic's model, this would stabilise the hybridisation via the Rashba coupling, pushing the Fermi surface reconstruction transition to higher temperatures, consistent with our observation that the crossover from insulating to metallic behaviour at 14 T is $\sim$70 K.
On the other hand, the resistance decreases with increasing field at low temperatures where we might expect the system to be driven more insulating via the polarisation of moments along the $b$-axis. Although, this could be explained by a reduction in scattering of electrons by spin fluctuations as the moments become polarised at high fields.

\section{Conclusion}
\label{sec:level6}
We have presented a comprehensive investigation  of the magnetic phase diagram of  Ca3Ru2O7 via magnetisation and elastic neutron scattering measurements. The previously observed IC state is stabilised between the \AFMa\ and \AFMb\ phases in the range 46.7 K to 49.0 K. With the application of the magnetic field along the $b$-axis, the IC broadens in temperature to between 35 K and 54 K at 5 T before collapsing via a phase transition to a field polarised state above 5.8 T. The data establishes the microscopic origin of the \textit{metamagnetic texture} found in ref~\cite{Sokolov2019}: frustration between commensurate states leading to DM-stabilised cycloidal order. Our phase diagram differs significantly from some published work probably due the use of twinned samples in their studies\cite{Bao2008}. We note that the origin of the discrepancy in the size of temperature-field region of the IC phase between our work and others remains unresolved\cite{Kikugawa2021}. We clearly observed the IC phase at all fields below 5 T, unlike previous work where it is only stabilised above 2 T. Work is continuing to resolve this issue. Furthermore, the microscope origin of the SRT remains elusive and needs further experimental and theoretical studies.

\section*{Acknowledgements}

The authors would like to thank Adam Walker, Michal Kwasigroch, Frank Kruger and Andrew Green for fruitful discussions. We also thank Pascal Fouilloux for technical support during the neutron diffraction experiment on D23. We gratefully acknowledge the Fédération Française de la Neutronique (2FDN) and Institut Laue Langevin (ILL) for beam time access to the neutron diffractometer D23.  Work at UCL is supported by the UK Engineering and Physical Sciences Research Council (Grants No. EP/N027671/1 and No. EP/N034694/1). C.D.D. was supported by the Engineering and Physical Sciences Research Council (EPSRC) Centre for Doctoral Training in the Advanced Characterisation of Materials under Grant No. EP/L015277/1.

\section*{Appendix A: Magnetic structure analysis}
\label{AppendixA}

\begin{table*}[!ht]
\centering
\begin{tabular}{|c|c|wc{1cm}wc{1cm}wc{1cm}|wc{1cm}wc{1cm}wc{1cm}|wc{1cm}wc{1cm}wc{1cm}|wc{1cm}wc{1cm}wc{1cm}|}
\hline
\hline
 &
   &
  \multicolumn{3}{|c|}{Equivalent position 1} &
  \multicolumn{3}{|c|}{Equivalent position 2} &
  \multicolumn{3}{|c|}{Equivalent position 3} &
  \multicolumn{3}{|c|}{Equivalent position 4} \\
 &
   &
  \multicolumn{3}{|c|}{x, y, z} &
  \multicolumn{3}{|c|}{-x,y+1/2,-z} &
  \multicolumn{3}{|c|}{x,y,-z} &
  \multicolumn{3}{|c|}{-x,y+1/2,z} \\
  \hline
IR  & Basis vector & $m_x$ & $m_y$ & $m_z$ & $m_x$ & $m_y$ & $m_z$ & $m_x$ & $m_y$ & $m_z$ & $m_x$ & $m_y$ & $m_z$ \\
\hline
\textit{$mY$}1 & $\Psi$1           & 1  & 0  & 0  & -1 & 0  & 0  & -1 & 0  & 0  & 1  & 0  & 0  \\
               & $\Psi$2           & 0  & 1  & 0  & 0  & 1  & 0  & 0  & -1 & 0  & 0  & -1 & 0  \\
               & $\Psi$3           & 0  & 0  & 1  & 0  & 0  & -1 & 0  & 0  & 1  & 0  & 0  & -1 \\
\hline
\textit{$mY$}2 & $\Psi$4           & 1  & 0  & 0  & -1 & 0  & 0  & 1  & 0  & 0  & -1 & 0  & 0  \\
               & $\Psi$5           & 0  & 1  & 0  & 0  & 1  & 0  & 0  & 1  & 0  & 0  & 1  & 0  \\
               & $\Psi$6           & 0  & 0  & 1  & 0  & 0  & -1 & 0  & 0  & -1 & 0  & 0  & 1  \\
\hline
\textit{$mY$}3 & $\Psi$7           & 1  & 0  & 0  & 1  & 0  & 0  & -1 & 0  & 0  & -1 & 0  & 0  \\
               & $\Psi$8           & 0  & 1  & 0  & 0  & -1 & 0  & 0  & -1 & 0  & 0  & 1  & 0  \\
               & $\Psi$9           & 0  & 0  & 1  & 0  & 0  & 1  & 0  & 0  & 1  & 0  & 0  & 1  \\
\hline
\textit{$mY$}4 & $\Psi$10          & 1  & 0  & 0  & 1  & 0  & 0  & 1  & 0  & 0  & 1  & 0  & 0  \\
               & $\Psi$11          & 0  & 1  & 0  & 0  & -1 & 0  & 0  & 1  & 0  & 0  & -1 & 0  \\
               & $\Psi$12          & 0  & 0  & 1  & 0  & 0  & 1  & 0  & 0  & -1 & 0  & 0  & -1 \\
\hline
\hline
\end{tabular}
\caption{Nonzero IR’s and associated basis vectors $\Psi$ for the space group Bb21m with kY = [1 0 0] and the Ru magnetic atoms at 8b site.  The 8b site consist of 8 crystallographically equivalent positions 1 : (x,y,z), 2 : (-x,y+1/2,-z),3 : (x,y,-z), 4 : (-x,y+1/2,z), 5 : (x+1/2,y,z+1/2), 6 : (-x+1/2,y+1/2,-z+1/2), 7 : (x+1/2,y,-z+1/2), 8: (-x+1/2,y+1/2,z+1/2). The B centering is broken and the magnetic moment of an atom in position 5-8 are antiparallel as in positions 1-4.}
\label{tableApp1}
\end{table*}

\begin{figure}[!h]
\centering{\includegraphics[width=0.8\linewidth]{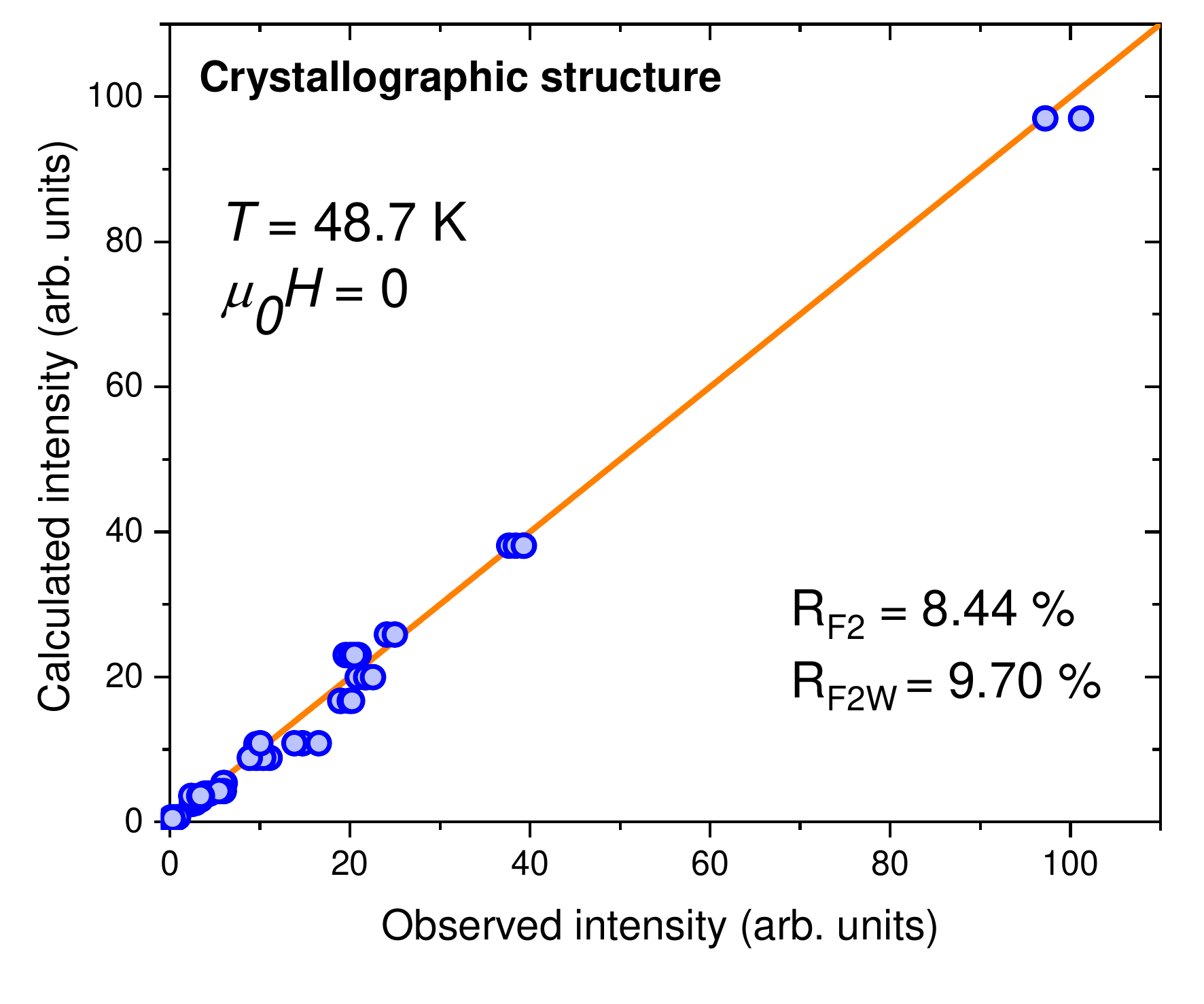}}
\caption{Representation of the crystallographic structure refinement performed with 70 magnetic reflections (21 independent ones) collected at zero field and $T = 48.7$~K leading to the following reliability factors: $R_{\rm{F2}}=8.44 \%$, $R_{\rm{F2w}}= 9.70~\%$, $R_{\rm{F}}=6.54~\%$ , and $\chi^2 = 23.6~\%$.}
\label{FigureSupMat5}
\end{figure}

In \caruo, the paramagnetic crystal structure is described in the space group Bb2$_1$m1’ (n$^{\circ}$36, non-standard setting). The Ru$^{4+}$ magnetic cation is in the general 8b Wyckoff position. As previously mentionned in the main text, we first began by refining the crystallographic structure by fixing all atomic coordinates with published values found in Ref.~\cite{Yoshida2005}. Fig.~\ref{FigureSupMat5} shows the Ritvelt refinement of the crystallographic structure at $T = 48.7$~K and $\mu_0H = 0$.

From the symmetry analysis point of view, we can distinguish two different cases depending on the nature of the observed propagation vector(s):
\begin{itemize}
    \item i: Commensurate antiferromagnetic phases with a Y-point $\bm{k}Y$= (1,0,0) propagation vector
    \item ii: Incommensurate antiferromagnetic phases with a propagation vector of the form $\bm{k}_{\Delta}$=($k_x$,0,0) , $k_x$ being irrational, part of the $\Delta$-line of the first Brillouin zone.
\end{itemize}

For each case, the decomposition of the magnetic representation in terms of the non-zero irreducible representations (IRs) for Ru site was examined using Basireps~\cite{Carvajal1993}, and their associated basis vectors are given: in the table~\ref{tableApp1} for kY and table~\ref{tableSupmat2} for $\bm{k}_{\Delta}$.


\begin{table}[]
\centering
\begin{tabular}{|c|c|wc{0.6cm}wc{0.6cm}wc{0.6cm}|wc{0.6cm}wc{0.6cm}wc{0.6cm}|}
\hline
\hline
    &              & \multicolumn{3}{|c|}{Equivalent position 1} & \multicolumn{3}{|c|}{Equivalent position 2} \\
    &              & \multicolumn{3}{|c|}{x, y, z}               & \multicolumn{3}{|c|}{x, y, -z}              \\
    \hline
IR  & Basis vector & $m_x$          & $m_y$           & $m_z$          & $m_x$           & $m_y$           & $m_z$          \\
\hline
$m\Delta$1 & $\Psi$1           & 1            & 0            & 0           & -1           & 0            & 0           \\
           & $\Psi$2           & 0            & 1            & 0           & 0            & -1           & 0           \\
           & $\Psi$3           & 0            & 0            & 1           & 0            & 0            & 1           \\
           \hline
$m\Delta$2 & $\Psi$4           & 1            & 0            & 0           & 1            & 0            & 0           \\
           & $\Psi$5           & 0            & 1            & 0           & 0            & 1            & 0           \\
           & $\Psi$6           & 0            & 0            & 1           & 0            & 0            & -1          \\
           \hline
           \hline
\end{tabular}
\caption{Nonzero IRs and associated basis vectors $\Psi$ for the space group Bb21m with $\bm{k}_{\Delta}$=($k_x,0,0)$. The magnetic atoms Ru at 8b site are split into two independent orbits: Ru$_1$ (0.2541 0.7503 0.4011) and Ru$_2$ (0.7459 0.2503 0.5989). For both orbit the same magnetic decomposition of the magnetic representation into two magnetic representation each having three basis vectors applies.}
\label{tableSupmat2}
\end{table}

This representational analysis approach was combined with magnetic space group determination using group theory program from ISODISTORT and Bilbao Crystallographic Server~\cite{Campbell2006, Aroyo2011}. In the kY case, the magnetic unit cell correspond to the crystalline unit cell and group theory predicts 4 possible maximum Magnetic Space Group (MSG, BNS notation): P$_{\rm{C}}$mc2$_1$ ($\sharp$~26.76), P$_{\rm{C}}$na2$_1$ ($\sharp$~33.154), P$_{\rm{C}}$mn2$_1$ ($\sharp$~31.133), P$_{\rm{C}}$ca2$_1$ ($\sharp$~29.109). There is a direct one to one correspondence of these MSG and the 4 possible Irreps given by representational analysis ($mY$1, $mY$2, $mY$3 and $mY$4 respectively).
In the second case (ii) where the magnetic structure is described by a single propagation vector belonging to the $\Delta$-line of the Brillouin zone (form $\bm{k}_{\Delta}$=(ky,0,0), there are two possible maximum Magnetic Super Space Group (MSSG): A2$_1$ma.1'(0,0,g)000s (corresponding to irrep $m\Delta$1) and A2$_1$ma.1'(0,0,g)0s0s (corresponding to irrep $m\Delta$2).

Commensurate antiferromagnetic structures:  the magnetic structure belongs to the Y-point symmetry ($\bm{k}$= (1,0,0). The irreducible magnetic representation and their associated three basis vectors are given in table~\ref{tableApp1}.

\begin{figure}[!h]
\centering{\includegraphics[width=\linewidth]{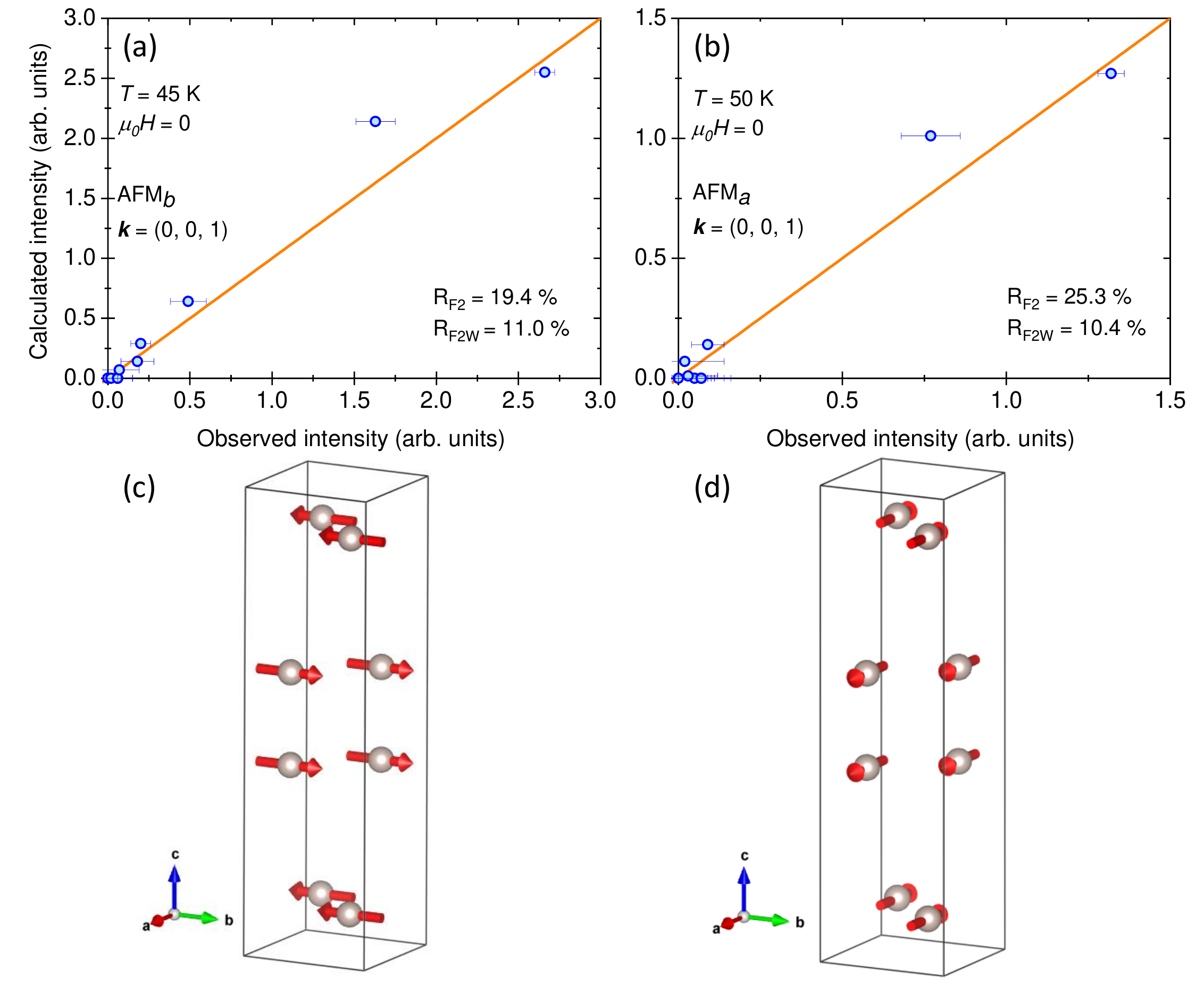}}
\caption{(a-b) Representation of the magnetic structure refinements at zero field for the \AFMb\ and \AFMa\ phases with propagation vector $\bm{k} = (0, 0, 1)$ performed with 28 magnetic reflections (10 independent ones) collected at zero field and $T = 45$~K and $T = 50$~K respectively. (c-d) Corresponding \AFMb\ and \AFMa\ magnetic structures.}
\label{FigureSupMat1}
\end{figure}

Table~\ref{tableSupMat3} along with Fig.~\ref{FigureSupMat1}(a-b) show the results of the magnetic structure refinements of the \AFMb\ and \AFMa\ performed at zero field and for both temperatures $T = 45$~K and $T = 50$~K respectively.  Fig~\ref{FigureSupMat1}(c-d) show the corresponding magnetic structures which consists of antiferromagnetically coupled ferromagnetic bilayers aligned along $b$ and $a$ respectively.

$\mu_0H = 0$: all irreps allowed by symmetry were tested and $mY$2 unambiguously leads to the best refinement with the magnetic moments aligned along the $b$-direction (see table~\ref{tableSupMat3}). The magnetic space group describing the magnetic structure is P$_{\rm{C}}$na2$_1$ ($\sharp$ 33.154). The magnetic arrangement consists of ferromagnetic double layers aligned along $b$, coupled antiferromagnetically (see Fig.~\ref{FigureSupMat1}(c)).
$T = 50$~K and $\mu_0H = 0$: all irreps allowed by symmetry were tested and $mY$4 unambiguously leads to the best refinement with the magnetic moments aligned along the a-direction (see table~\ref{tableSupMat3}). The magnetic space group describing the magnetic structure is P$_{\rm{C}}$ca2$_1$ ($\sharp$ 29.109). The magnetic arrangement consists of ferromagnetic double layers aligned along $a$, coupled antiferromagnetically (see Fig.~\ref{FigureSupMat1}(d)).

\begin{table}[]
\centering
\begin{tabular}{|P{2cm}|P{1cm}|P{2.5cm}|P{2.5cm}|}
\hline
\hline
Temperature-field     &        & 45 K - 0 T   & 50 K – 0 T     \\
\hline
Magnetic structure &  & Antiferromagnetic structure “\AFMb” & Antiferromagnetic structure “\AFMa” \\
\hline
MSSG                  &        & P$_{\rm{C}}$na2$_1$         & P$_{\rm{C}}$ca2$_1$         \\
IR                    &        & $mY$2            & $mY$4            \\
Propagation vector    &        & (0, 0, 1)  & (0, 0, 1)  \\
Ru                    & $M$($\mu_{\rm{B}}$) & 1.533(29)      & 1.045(19)      \\
                      & $m_x$     & 0              & 1.045(19)      \\
                      & $m_y$     & 1.533(29)      & 0              \\
                      & $m_z$     & 0              & 0              \\
Number of Reflections &        & 28 (10 independents)             & 28 (10 independents)    \\
$R_{\rm{F2}}$         &        & 19.5           & 25.4           \\
$R_{\rm{F2w}}$        &        & 11.1           & 10.4           \\
$R_{\rm{F}}$          &        & 19.4           & 36.5           \\
$\chi^2$              &        & 3              & 1.42           \\
\hline
\hline
\end{tabular}
\caption{Refined parameters of \caruo\ commensurate antiferromagnetic structures at 45K and 49K under zero field. Both structures have a propagation vector of the form $kY$ = (1, 0, 0) (see table~\ref{tableApp1} for representational analysis).  $M$ corresponds to the modulus of the magnetic moment. $R_{\rm{F2}}$, $R_{\rm{F2w}}$, $R_{\rm{F}}$ , and $\chi^2$ are the reliability factors.}
\label{tableSupMat3}
\end{table}

\begin{table}[]
\centering
\begin{tabular}{|P{3cm}|P{1.5cm}|P{4cm}|}
\hline
\hline
Temperature-field       &         & 48.7~K – 0~T                                           \\
\hline
Magnetic structure &         & Incommensurate elliptical cycloid  \\
\hline
MSSG                    &         & A2$_1$ma.1'(0,0,g)0s0s       \\
IR                      &         & $m\Delta$2                   \\
Propagation vector      &         & ($\delta$, 0, 1)                 \\
$k_y$                   &         & 0.975(5)                      \\
Ru                      & $m_{a}^j$ ($\mu_B$) & 0.83(1)            \\
                        & $m_{b}^j$ ($\mu_B$) & 1.30(1)             \\
$\Delta \Phi$ (unit of 2$\pi$)    &         & 0                    \\
Number of Reflections   &         & 59 (15 independents)            \\
$R_{\rm{F2}}$           &         & $27.3 \%$                       \\
$R_{\rm{F2w}}$          &         & $12.1 \%$                       \\
$R_{\rm{F}}$            &         & $47.9 \%$                       \\
$\chi^2$                &         & $0.622 \%$                      \\
\hline
\hline
\end{tabular}
\caption{Refined parameters of \caruo\ incommensurate cycloid magnetic structures at 48.7~K under zero magnetic field. $\Delta \Phi$ corresponds to the magnetic phase difference between Ru$_1$ (0.2541 0.7503 0.4011) and Ru$_2$ (0.7459 0.2503 0.5989). The refined amplitude of the magnetic moment components $m_{a}^j$ and $m_{b}^j$ (along $\bm{a}$ and $\bm{b}$ respectively) are given. R$_{\rm{F2}}$, $R_{\rm{F2w}}$, $R_{\rm{F}}$, and $\chi^2$ are the reliability factors.}
\label{table1}
\end{table}

Determination of incommensurate magnetic structures with a propagation vector of the form $\bm{k}_{\Delta}$= ($k_x$,0,0), $k_x$ being irrational, part of the $\Delta$-line of the first Brillouin zone: The representational analysis tells us that the Ru site (8b Wyckoff position) splits into two orbits: Ru$_1$ (0.2541 0.7503 0.4011) and Ru$_2$ (0.7459 0.2503 0.5989). For each orbit, the same decomposition of the magnetic representation into two magnetic representation applies. The irreducible magnetic representation and their associated three basis vectors are given in table~\ref{tableSupmat2}. The two irreps allowed by symmetry were tested and $m\Delta$2 unambiguously leads to the best refinement. The two orbits Ru$_1$ and Ru$_2$ were constrained to have the same magnetic moment and only the magnetic phase difference $\Delta \Phi$ between these two sites was refined. The magnetic super space group describing the magnetic structure is A2$_1$ma.1'(0,0,g)0s0s. The refined magnetic structure is an elliptical cycloid propagating along $a$-direction. The layers of ferromagnetically coupled magnetic moments turn in the $(a, b)$ plane and are arranged antiferromagnetically along $c$. Table~\ref{table1} summarizes the results of the magnetic refinement.

\begin{figure}[!h]
\centering{\includegraphics[width=\linewidth]{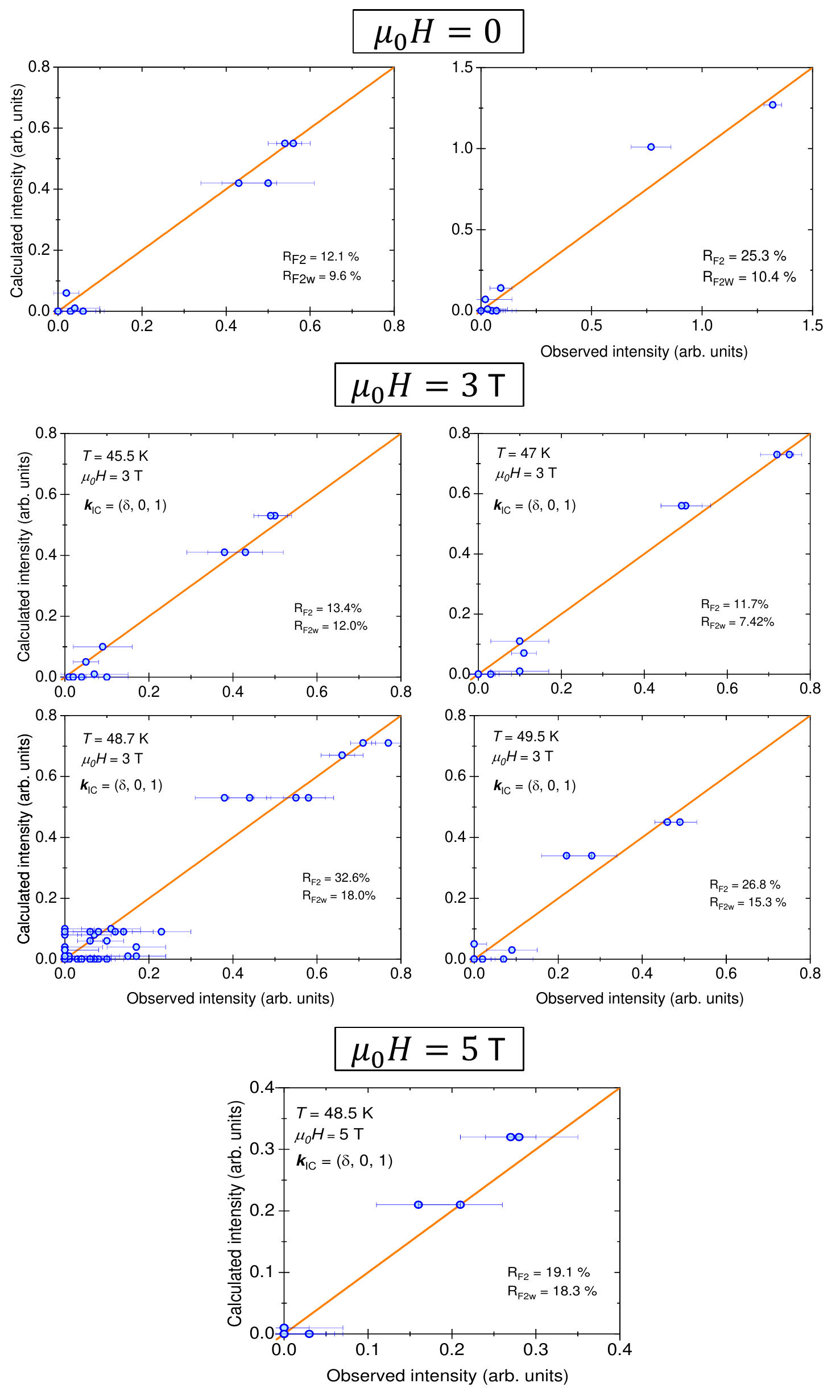}}
\caption{Representation of the magnetic structure refinements of the IC phase for $\mu_0H = 0, 3$ and $5$~T and different temperatures.}
\label{FigureSupMat2}
\end{figure}

Fig.~\ref{FigureSupMat2} shows the numerous magnetic structure refinements for different values of magnetic field and temperature. Those results allowed to extract the field-temperature evolution of the IC cycloid phase depicted in Fig.~\ref{Figure4}(b-c).

\section*{Appendix B: $H-T$ phase diagram with $\vec{H} || \vec{a}$}
\label{AppendixB}

\begin{figure}[!h]
\centering{\includegraphics[width=\linewidth]{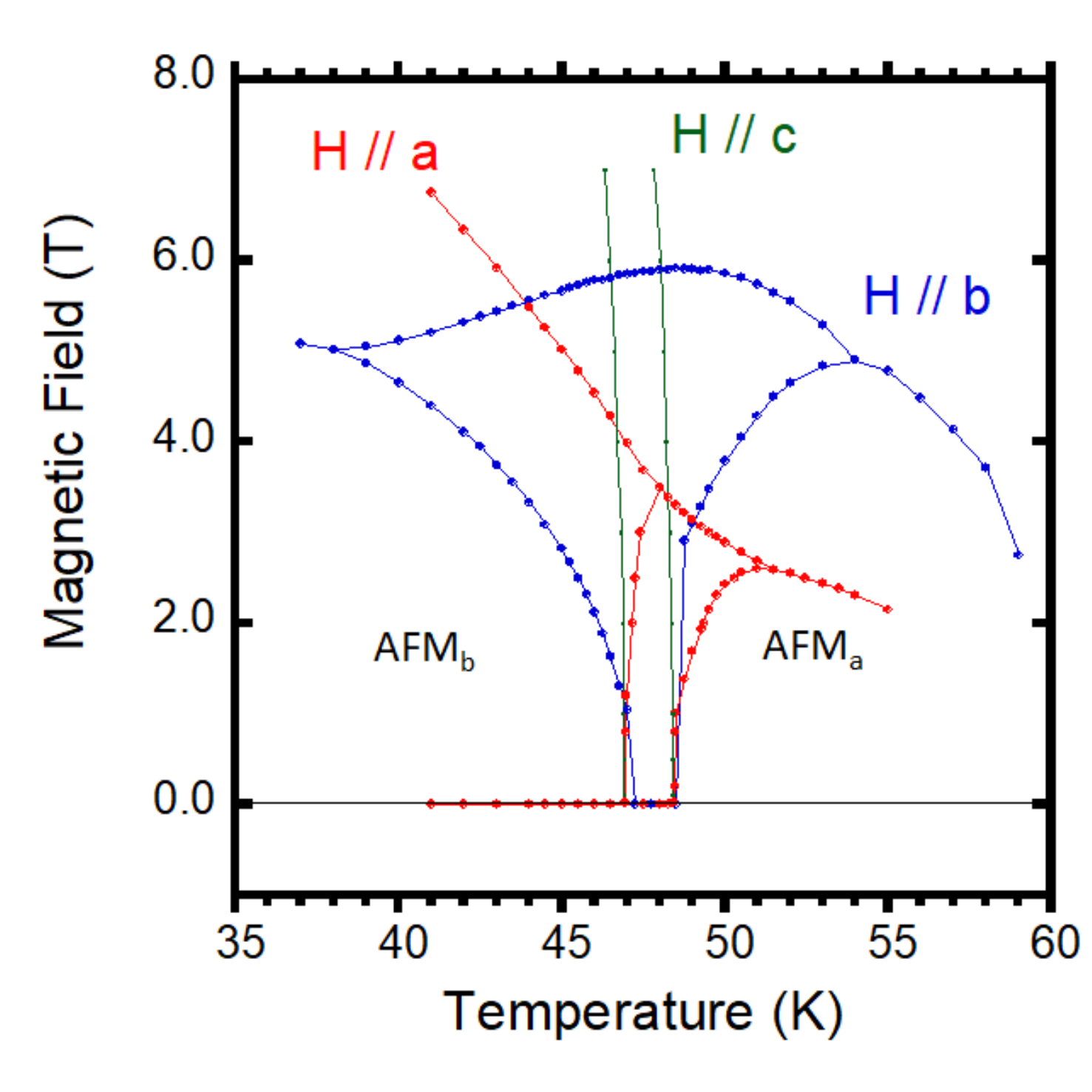}}
\caption{$H-T$ phase diagrams for $\bm{H} \parallel \bm{a}$ (red points), $\bm{H} \parallel \bm{b}$ (blue points) and $\bm{H} \parallel \bm{c}$ (green points). Lines are guide to the eye.}
\label{FigureSupMat3}
\end{figure}

To check our hypothesis about the twinned crystals used in Ref.~\cite{Bao2008}, we probed the $H-T$ phase diagram with $\bm{H} \parallel \bm{a}$. For this purpose, We performed magnetometry measurements similar to Sec.~\ref{subsec: 3: 1} with $\bm{H} \parallel \bm{a}$. Fig.~\ref{FigureSupMat3} shows the superposition of phase diagrams obtained with $\bm{H} \parallel \bm{a}$ and $\bm{H} \parallel \bm{b}$. The steep phase transition line around $T \simeq 48$~K and $\mu_0 H \simeq 4$~T in the phase diagram with $\bm{H} \parallel \bm{a}$ is remarkably similar to Bao's leading us to conclude that their crystals were twinned. For completeness, the phase diagram with $\bm{H} \parallel \bm{c}$ was also probed but no effect onto the phase boundaries has been observed.

\bibliographystyle{apsrev4-2}
\bibliography{biblio}

\end{document}